\documentclass[a4paper,12pt]{article}
\pdfoutput=1 

\usepackage{jheppub} 

\usepackage[T1]{fontenc} 
\usepackage{lmodern}
\usepackage{tikz}

\let\oldFootnote\footnote
\newcommand\nextToken\relax

\renewcommand\footnote[1]{%
    \oldFootnote{#1}\futurelet\nextToken\isFootnote}

\newcommand\isFootnote{%
    \ifx\footnote\nextToken\textsuperscript{,}\fi}

\usepackage{enumerate}

\def\AdS{{\bf AdS}}

\def\id{{1 \kern-.28em {\rm l}}}

\def\K3{{\bf K3}}
\def\journal#1&#2(#3){\unskip, \sl #1\ \bf #2 \rm(19#3) }
\def\andjournal#1&#2(#3){\sl #1~\bf #2 \rm (19#3) }

\def\bar{\overline}
\def\hat{\widehat}
\def\ie{{\it i.e.}}
\def\eg{{\it e.g.}}

\def\tilde{\widetilde}

\def\frac#1#2{{#1\over#2}}

\def\half{\frac12}

\def\inbar{\,\vrule height1.5ex width.4pt depth0pt}
\def\IC{\relax\hbox{$\inbar\kern-.3em{\rm C}$}}
\def\IR{\relax{\rm I\kern-.18em R}}
\def\IP{\relax{\rm I\kern-.18em P}}

\catcode`\@=11
\def\slash#1{\mathord{\mathpalette\c@ncel{#1}}}
\def\MM{{\cal M}}
\def\NN{{\cal N}}

\def\underrel#1\over#2{\mathrel{\mathop{\kern\z@#1}\limits_{#2}}}

\catcode`\@=12

\def \sinh{{\rm sinh}}

\def\ie{{\it i.e.}}
\def\eg{{\it e.g.}}

\title{Entanglement Beyond $\AdS$}

\author{Soumangsu Chakraborty$^a$,}
\author{Amit Giveon$^a$,}
\author{Nissan Itzhaki$^b$}
\author{and David Kutasov$^c$}

\affiliation[a]{Racah Institute of Physics,\\The Hebrew University, Jerusalem 91904 Israel}
\affiliation[b]{Physics Department,\\ Tel-Aviv University, Israel Ramat-Aviv, 69978, Israel}
\affiliation[c]{EFI and Department of Physics,\\ University of
Chicago, 5640 S. Ellis Av., Chicago, IL 60637, USA}

\abstract{We continue our study of string theory in a background that interpolates between $AdS_3$ in the infrared and a linear dilaton spacetime $\IR^{1,1}\times\IR_\phi$ in the UV. This background corresponds via holography to a $CFT_2$ deformed by an operator of dimension $(2,2)$. We discuss the structure of spatial entanglement in this model, and compare it to the closely related $T\bar T$ deformed $CFT_2$.}

\begin{document}
\maketitle
\flushbottom

\section{Introduction}
We have recently \cite{Giveon:2017nie,Giveon:2017myj,Asrat:2017tzd,Chakraborty:2018aji} initiated a detailed study of string theory in backgrounds of the form 
\begin{eqnarray}\label{theback}
\MM_3\times \NN,
\end{eqnarray}
where $\NN$ is a compact space, and $\MM_3$ interpolates between a three dimensional linear dilaton background in the ultraviolet (UV) and $AdS_3$ in the infrared (IR). 

From the UV point of view, these backgrounds can be interpreted as the bulk description of certain two dimensional vacua of Little String Theory (LST) that contain $N\gg1$ fundamental strings \cite{Aharony:1998ub,Giveon:1999zm}. From this point of view the high energy physics is that of the underlying LST, while at low energies one approaches a CFT  that is dual to the $AdS_3$ background corresponding to the near-horizon geometry of strings in the linear dilaton background.  

From the IR point of view, these backgrounds can be thought of as  irrelevant deformations of the two dimensional conformal field theories $(CFT_2)$ dual to the above $AdS_3$ backgrounds by a certain universal quasi-primary operator of dimension $(2,2)$,  \cite{Giveon:2017nie,Giveon:2017myj,Asrat:2017tzd,Chakraborty:2018aji}. Superficially, one might think that such a description would be incomplete, since it involves flowing up the renormalization group (RG), a process that is usually ambiguous, however, we presented evidence (both worldsheet and spacetime) that in this case the situation is much better, and one can in fact flow up the RG. 

In this note we will discuss the structure of spatial entanglement in these models, between an interval of length $L$ and its complement. Since the models are non-local at short distances, we expect their entanglement entropy $S(L)$ to exhibit non field-theoretic behavior at small $L$. One of our main goals is to investigate this non-locality. In particular, we will be interested in the question what happens when the size of the interval, $L$, approaches the non-locality scale of the model. 

We will also discuss the dependence of the entanglement entropy on the UV cutoff. In two dimensional QFT it is well known that $S(L)$ depends logarithmically on the cutoff, but the quantity 
\begin{eqnarray}\label{ccclll}
C(L)=3L{\partial S(L)\over\partial L}
\end{eqnarray}
is finite \cite{Casini:2006es}. At RG fixed points it approaches a constant, equal to the Virasoro central charge. As $L$ varies between $0$ and $\infty$, the entropic $c$-function $C(L)$ \eqref{ccclll}  interpolates (monotonically) between the UV and IR central charges, respectively. It is natural to ask what is the cutoff dependence of $C(L)$ in the non-local models of \cite{Giveon:2017nie,Giveon:2017myj,Asrat:2017tzd,Chakraborty:2018aji}. 
  
As explained in \cite{Giveon:2017nie,Giveon:2017myj,Asrat:2017tzd,Chakraborty:2018aji}, the string theory models we will discuss are closely related to $T\bar T$ deformed $CFT_2$, a subject that received some attention recently following the work of \cite{Smirnov:2016lqw,Cavaglia:2016oda}; see \eg\ \cite{McGough:2016lol,Shyam:2017znq,Giribet:2017imm,Kraus:2018xrn,Cardy:2018sdv,Cottrell:2018skz,Aharony:2018vux,Dubovsky:2018dlk,Bonelli:2018kik}, and \cite{Guica:2017lia,Bzowski:2018pcy,Sakamoto:2018krs} for work on related systems. It is natural to ask whether the general form of the entanglement entropy in these models is similar to that found in the string backgrounds we will study. Unfortunately, it is difficult to perform the full calculation in the field theory models of \cite{Smirnov:2016lqw,Cavaglia:2016oda}, but we will show that perturbatively in the coupling one finds a similar structure in the two cases.

The plan of this paper is the following. In section \ref{sec2} we briefly review some relevant aspects of the construction of \cite{Giveon:2017nie,Giveon:2017myj,Asrat:2017tzd,Chakraborty:2018aji}, and of previous work on entanglement entropy \cite{Calabrese:2009qy,Nishioka:2009un}. In section \ref{sec3} we compute the holographic entanglement entropy and $c$-function \eqref{ccclll} of a line segment of length $L$ in the models of \cite{Giveon:2017nie,Giveon:2017myj,Asrat:2017tzd,Chakraborty:2018aji}. We work out their behavior at large $L$ (for the purpose of comparing to the field theory calculation), and near a critical value $L_{\rm min}$, where the $c$-function diverges.

In section \ref{sec4} we discuss the calculation of entanglement entropy in field theory. We take as an example the theory of $N$ complex fermions, deform it by the operator $\lambda T\bar T$, and compute the entanglement entropy to first non-trivial order in $\lambda$. We find that the result is very similar to the holographic calculation. In section \ref{sec5} we generalize the holographic calculation to finite temperature. In section \ref{sec6} we comment on our results and discuss possible avenues for further work.

\section{Review}\label{sec2}

\subsection{An irrelevant deformation of $AdS_3/CFT_2$}\label{sec2.1}

The work of \cite{Giveon:2017nie,Giveon:2017myj,Asrat:2017tzd,Chakraborty:2018aji} was inspired by the observation \cite{Smirnov:2016lqw,Cavaglia:2016oda} that a particular irrelevant deformation of a generic $CFT_2$ by an operator bilinear in stress tensors leads to a theory with some interesting properties. 

This theory appears to be well defined at high energies, despite the fact that it involves a flow up the renormalization group. Moreover, it seems to be, at least to some extent, solvable. For example, the authors of \cite{Smirnov:2016lqw,Cavaglia:2016oda} were able to compute the spectrum of the theory exactly. The density of states interpolates between that of a $CFT_2$ at low energies and one with Hagedorn growth at high energies. Thus, it is not a standard local QFT, in the sense that the high energy behavior is not governed by a UV fixed point. In supersymmetric models, the deformation preserves supersymmetry, since the deforming operator is a top component of a superfield.

If the original $CFT_2$ has a holographic $(AdS_3)$ dual, the theory of \cite{Smirnov:2016lqw,Cavaglia:2016oda} corresponds to a double trace deformation of the original duality.\footnote{See \eg\ \cite{Witten:2001ua,Berkooz:2002ug} for discussions of double trace deformations.}  The main observation of \cite{Giveon:2017nie}  was that there is a single trace deformation of the $CFT_2$, that shares many of the properties mentioned above. In particular, it is universal, in the sense that any $CFT_2$ that has an $AdS_3$ dual (at least one supported by $NS$ $B$-field flux) contains this deformation. It is under control despite being irrelevant since it corresponds to a truly marginal deformation of the worldsheet theory, 
\begin{eqnarray}\label{JJbar}
\delta\mathcal{L}_{ws}=\lambda J^-\bar{J}^-,
\end{eqnarray}
where $J^-$ is the holomorphic worldsheet $SL(2,\IR)$ current whose zero mode gives rise to the boundary Virasoro generator $L_{-1}$, and similarly for $\bar J^-$, $\bar L_{-1}$ \cite{Giveon:1998ns}. Deforming the worldsheet theory by \eqref{JJbar} corresponds to deforming the boundary $CFT_2$ by 
\begin{eqnarray}\label{D(x)}
\delta\mathcal{L}=\tilde\lambda D(x),
\end{eqnarray}
where $D(x)$ is a certain dimension $(2,2)$ quasi-primary constructed in \cite{Kutasov:1999xu}, and $\tilde\lambda$ is proportional to $\lambda$ \eqref{JJbar}.  

The deformation \eqref{JJbar}, \eqref{D(x)} corresponds in the bulk to a deformation of the metric, dilaton and $B$ field, from $AdS_3$ to a background that we will refer to as $\mathcal{M}_3$ (see \eqref{theback}). It is described by \cite{Forste:1994wp}
\begin{eqnarray}
ds^2&=&f^{-1}\left(-dt^2+dx^2\right)+k\alpha'\frac{dU^2}{U^2},\nonumber \\
e^{2\Phi}&=&\frac{g_s^2}{kU^2}f^{-1},\label{background}\\
dB&=&\frac{2i}{U^2}f^{-1}\epsilon_3,\nonumber
\end{eqnarray}
where $f=1+\frac{1}{kU^2}$, and $k$ is the level of the worldsheet $SL(2,\IR)$ current algebra of the model. 

The background \eqref{background} interpolates between $AdS_3$ (with string coupling $g_s$) in the infrared region $U\to 0$, and a linear dilaton spacetime 
 \begin{eqnarray}\label{UV geometry}
 \IR^{1,1}\times \IR_\phi
\end{eqnarray}
(with $\phi\sim\ln U$) in the UV region $U\to\infty$. The scale at which the geometry makes the transition between the two depends on $\lambda$ \eqref{JJbar}; it has been set to a convenient value in \eqref{background}, without loss of generality (see \cite{Giveon:2017nie,Giveon:2017myj} for a more detailed discussion of this and related issues). 

The UV geometry \eqref{UV geometry} describes a two dimensional vacuum of Little String Theory \cite{Aharony:1998ub}. The high energy density of states exhibits Hagedorn growth with inverse Hagedorn temperature (see \eg\ \cite{Aharony:2004xn} for a review)
\begin{eqnarray}\label{H temp}
\beta_H=2\pi\sqrt{k\alpha'}~.
\end{eqnarray}
One can think of the full geometry \eqref{background} as obtained from \eqref{UV geometry} by adding $N$ fundamental strings filling $\IR^{1,1}$   \cite{Giveon:1999zm}. As one approaches the strings, the coupling stops growing, and saturates at a value $g_s^2\propto 1/N$. From this perspective it is clear that if the original LST is supersymmetric (in spacetime), the deformation preserves supersymmetry, since the strings preserve some of the supersymmetry of the original LST.\footnote{An example of this construction is the near-horizon geometry of $k$ $NS5$-branes wrapped around a compact four dimensional manifold, \eg\ $T^4$ or $K_3$, which gives rise to \eqref{UV geometry}. The background \eqref{background} is obtained by adding $N$ fundamental strings stretched along the remaining $\IR^{1,1}$.}

One can study the thermodynamics of the model \eqref{background} by analyzing black holes in $\mathcal{M}_3$. This gives \cite{Giveon:2017nie} an entropy that interpolates between a Cardy entropy with $c=6kN$ at low energies, and Hagedorn entropy $S=\beta_H E$ at high energies. 

As mentioned above, the single trace perturbation \eqref{JJbar}, \eqref{D(x)} is different from the double trace $T\bar T$ deformation, which is the direct analog of \cite{Smirnov:2016lqw,Cavaglia:2016oda}. A useful heuristic way of thinking about the difference between the two is the following. Suppose the boundary $CFT_2$ corresponding to a particular string theory on $AdS_3$ was a symmetric product $\MM^N/S_N$. Then one could consider two different deformations related to that of \cite{Smirnov:2016lqw,Cavaglia:2016oda} -- the $T\bar T$ deformation of the full symmetric product, and the $T\bar T$ deformation of the block $\MM$. The first can be thought of as a double trace deformation, while the second is a single trace one. This picture is heuristic since the boundary $CFT_2$ is not expected in general to be a symmetric product, though as discussed in \cite{Giveon:2017nie,Giveon:2017myj,Asrat:2017tzd} (following \cite{Argurio:2000tb,Giveon:2005mi}) there are many connections.

\subsection{Entanglement entropy}\label{sec2.2}

In this subsection we review the results on entanglement entropy in two dimensional quantum field theory that we will use below; see \eg\   \cite{Calabrese:2009qy,Nishioka:2009un} for more detailed discussions.

Consider a two dimensional QFT on $\IR^{1,1}$, in its vacuum state $|0\rangle$. The vacuum is a pure state; thus, the corresponding density matrix, $\rho=|0\rangle\langle0|$, satisfies $\rho^2=\rho$. We divide a spatial slice, $\IR$, into two regions: $A=I_L$, an interval of length $L$, and $B$, the complement of $A$. The quantity of interest is the  Von-Neumann entropy of the density matrix $\rho_A=\mathrm{Tr}_B\rho$, which can be thought of as the entropy seen in the pure state $|0\rangle$ by an observer in region $A$ that does not have access to region $B$.

One way to calculate it is to start with the $n^{\text{th}}$ Renyi entropy $R_n$ given by 
\begin{eqnarray}\label{renyi}
R_n=\frac{1}{1-n}\ln\mathrm{Tr}\rho_A^n,
\end{eqnarray}
continue the result to arbitrary (non-integer) $n$, and take the limit
\begin{eqnarray}\label{EntEntp}
S_{EE}=\lim_{n\to 1}R_n=-\frac{d}{dn}\mathrm{Tr}\rho_A^n|_{n=1}=-\rho_A\ln\mathrm{Tr}\rho_A.
\end{eqnarray}
This is known as the replica trick.

When the two dimensional QFT is conformal, the authors of  \cite{Calabrese:2009qy}  have shown that the Renyi entropy \eqref{renyi} can be computed by considering the product of $n$ copies of the CFT, constructing the lowest dimension $\mathbb{Z}_n$ twist field $S_n$ (where the $\mathbb{Z}_n$ cyclically permutes the $n$ copies), and computing the two point function
\begin{eqnarray}\label{trace}
\mathrm{Tr}\rho^n_A=\langle S_n(u)S_n(v)\rangle,
\end{eqnarray} 
where $u$ and $v$ are the endpoints of the interval $I_L$, with $L=|u-v|$. The twist field $S_n$ is a conformal primary of dimension\footnote{$c$ is the central charge of the original $CFT$.}
\begin{eqnarray}\label{dim twist}
\Delta(S_n)=\frac{c}{24}\left(n-\frac{1}{n}\right).
\end{eqnarray}
Its two point function is given by 
\begin{eqnarray}\label{twist ope}
\langle S_n(u)S_n(v)\rangle\sim \frac{1}{(u-v)^{4\Delta(S_n)}}~.
\end{eqnarray}
Plugging this into \eqref{renyi}, \eqref{EntEntp}, \eqref{trace} one finds the famous result \cite{Holzhey:1994we}
\begin{eqnarray}\label{see}
S_{EE}=\frac{c}{3}\log \frac{L}{a}.
\end{eqnarray}
As mentioned above, the dependence on the UV cutoff $a$ can be eliminated by considering the $c$-function $C(L)$ \eqref{ccclll}. 

The theories we will be studying here are not conformal, so we need the generalization of the results of \cite{Calabrese:2009qy} to the case of a general $QFT_2$. One can still use the construction of \cite{Calabrese:2009qy}, except now the correlation function \eqref{trace} is computed in the symmetric product of perturbed $CFT$'s.

In a $CFT_2$ with a gravity dual, one can calculate the  entanglement entropy \eqref{EntEntp} directly by using the results of Ryu and Takayanagi  \cite{Nishioka:2009un}. The entanglement entropy is proportional to the area of the minimal surface $\Sigma$ connecting the two points $u$ and $v$ on the boundary through the bulk at fixed time, and wrapping $\NN$ \eqref{theback}. 

This prescription is also valid for non-conformal boundary theories, whose gravity dual is not asymptotically $AdS_3$. For such theories the area functional that needs to be minimized is (see \eg\ \cite{Klebanov:2007ws}) 
\begin{eqnarray}
S_{EE}=\frac{1}{4G_N}\int_{\Sigma} d^8\sigma e^{-2(\Phi-\Phi_0)}\sqrt{G},\label{EE}
\end{eqnarray}
where $G_N$ is the 10 dimensional Newton constant, $\sigma$ are coordinates on the eight dimensional hypersurface $\Sigma$, $\Phi$ is the dilaton field and $G$ is the determinant of the induced string frame metric on $\Sigma$. Just like in the QFT calculation, the holographic entanglement entropy is, in general, UV divergent. To study the dependence on the UV cutoff, one can take the radial coordinate $U$ in \eqref{background} to run over the range $0\le U\le U_{\rm max}$, and investigate the dependence of the results on $U_{\rm max}$.

\section{Bulk calculation: zero temperature}\label{sec3}

\subsection{Holographic entanglement entropy}

In this section we calculate the entanglement entropy of string theory on $\MM_3\times \NN$  \eqref{theback}, \eqref{background}. As explained in the previous section, to do that we need to find a minimal surface connecting two points a distance $L$ apart on the boundary of $\MM_3$ (at fixed time), while wrapping all of $\NN$. One can think about this minimal surface as a configuration $U=U(x)$ which has the property that as $x$ approaches the endpoints of the interval, $x_{\rm end}=\pm L/2$, $U$ approaches the boundary, $U\to\infty$; between the two endpoints $U(x)$ decreases to some minimal value $U=U_0$ (which by symmetry is achieved at $x=0$) and then goes back up to infinity.

The area functional \eqref{EE} for $U(x)$ takes in this case the form 
\begin{eqnarray}\label{EE T=0}
S_{EE}&= &\frac{\sqrt{k}}{4G_N^{(3)}}\int_{-\frac{L}{2}}^{\frac{L}{2}} dx \sqrt{\frac{(kU^2+1)(U^4+\alpha'(kU^2+1)(\partial_x U)^2)}{U^2}}~.
\end{eqnarray} 
The equation of motion is 
\begin{eqnarray}\label{constraint}
\frac{U^3\sqrt{kU^2+1}}{\sqrt{U^4+\alpha'(kU^2+1)(\partial_x U)^2}}&=&U_0\sqrt{kU_0^2+1}~,
\end{eqnarray}
with the initial conditions $U(x=0)=U_0$ and $\partial_x U|_{x=0}=0$. As usual in such calculations (see \eg\ \cite{Klebanov:2007ws, Nishioka:2009un}), the size $L$ of the entangling region on the boundary, and the minimal value of $U$ along the curve $U(x)$, $U_0$, are related,  via
\begin{eqnarray}
L(U_0)=\frac{2\sqrt{\alpha'}}{U_0}\int_1^\infty dy \frac{\sqrt{kU_0^2y^2+1}}{y^2\sqrt{y^2\left(\frac{kU_0^2y^2+1}{kU_0^2+1}\right)-1}}~, \label{T=0 L}
\end{eqnarray}
where $y=U/U_0$. The integral \eqref{T=0 L} is convergent in the UV (\ie\ as $y\to\infty$), so we do not need to introduce a UV cutoff.
\begin{figure}[h]
\centering
    \includegraphics[width=.5\textwidth]{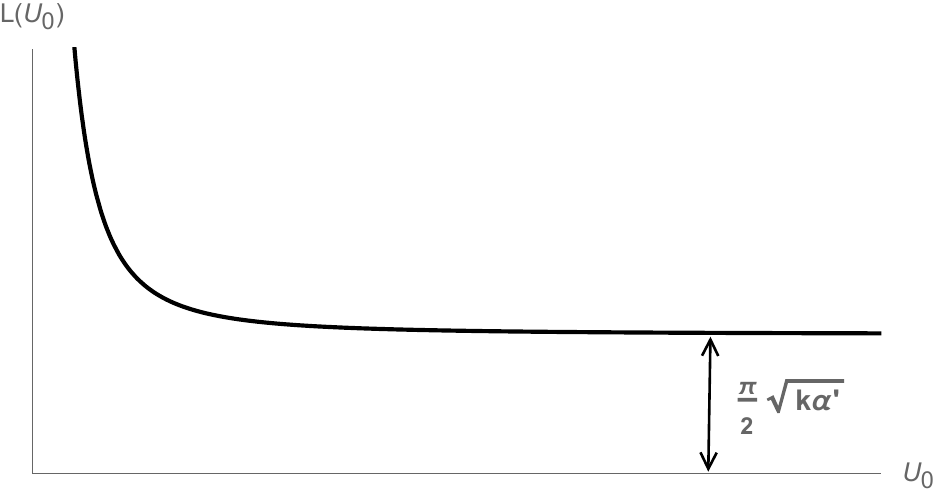}
    \caption{The size of the entangling region $L$ as a function of $U_0$ in $\mathcal{M}_3$.}
    \label{M3 T=0 EE L vs U0}
\end{figure} 

\begin{figure}
\begin{center}
\begin{tikzpicture}[scale=.6, transform shape]
  \draw [black,thick,domain=122:418] plot ({2.5*cos(\x)}, {1.5*sin(\x)});
  \draw [black,thick,domain=1.25:6] plot ({1.3}, {\x}); 
    \draw [black,thick,domain=1.25:6] plot ({-1.3}, {\x}); 
    \draw [black,thick,->,domain=-3:3] plot ({\x}, {6}); 
    \draw [dashed,black,thin,domain=-3:3] plot ({\x}, {5});
    \draw [dashed,black,thin,domain=-3:3] plot ({\x}, {1.25});
    \draw [dashed,black,thin,domain=-3:3] plot ({\x}, {-1.5});
      \draw [red,thin,domain=-2:2] plot ({.5*\x},{(1.325*\x)^2-1}); 
      \draw [blue,thin,domain=-1.42:1.42] plot ({.2*\x},{(\x)^2+4}); 
      \draw [green,dashed,thin,domain=-1.5:6.2] plot ({.2}, {\x});
       \draw [green,dashed,thin,domain=-1.5:6.2] plot ({-.2}, {\x});
    \draw [thick,->] (3.5,-1) -- (3.5,5);
    \draw (3.8,4.5) node {$U$};
      \draw (2.8,6.2) node {$x$};
      \draw (-3.8,6) node {$U=\infty$};
      \node[rotate=90] at (-1.8,3.5) {\text{Linear dilaton}};
      \node[rotate=90] at (-1.5,3.5) {\text{regime}};
      \node at (-1.3,-.4) {$AdS_3$};
      \node at (-1.3,-.7) {\text{regime}};
      \node at (-3.5,5) {$U_{\rm max}$};
           \draw (-3.8,-1.5) node {$U=0$};
           \draw [black,thick,<->,domain=-.2:.2] plot ({\x}, {6.2}); 
           \node at (0,6.5) {$L_{\rm min}$};
           \node at (-3.8,1.25) {$U\sim \frac{1}{\sqrt{k}}$};
\end{tikzpicture}
\caption{The background $\mathcal{M}_3$  gives rise at small $U$ to $AdS_3$, and at large $U$ to a linear dilaton background. The two are smoothly connected at $U\sim\frac{1}{\sqrt{k}}$. For large $L$ the bottom of the minimal  RT surface is deep inside the $AdS_3$ region (red curve), while for $L\sim L_{\rm min}$ it is in the linear dilaton region (blue curve). The properties of the corresponding entanglement entropy are accordingly different.}
  \label{spacetime}
  \end{center}
\end{figure}
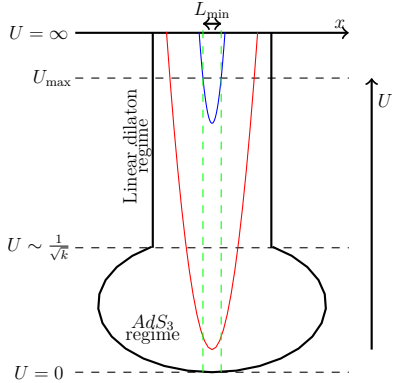

Figure (\ref{M3 T=0 EE L vs U0}) shows a numerical plot of $L$ as a function of $U_0$.  
As $L\to\infty$, $U_0\to 0$, so the bottom of the minimal surface is deep inside the $AdS_3$ region of the bulk geometry (red curve in figure (\ref{spacetime})). In this regime, the behavior of the entanglement entropy is dominated by the $AdS_3$ region, in agreement with standard RG intuition. 

As $L$ decreases, $U_0$ increases and ultimately diverges as $L$ approaches a minimal value $L_{\rm min}=\frac{\pi}{2}\sqrt{k\alpha'}$ (blue curve in figure (\ref{spacetime})). For $L\sim L_{\rm min}$,  the behavior of the entanglement entropy is dominated by the linear dilaton region of the geometry \eqref{background}. 

The entanglement entropy is obtained by plugging \eqref{constraint} in \eqref{EE T=0}. One finds
\begin{eqnarray}
S_{EE}&= & \frac{\sqrt{k\alpha'}}{2G_N^{(3)}}\int_{U_0}^{U_{\rm max}} \frac{dU}{U}\frac{(kU^2+1)}{\sqrt{1-\left(\frac{kU_0^2+1}{kU^2+1}\right)\frac{U_0^2}{U^2}}}~.\label{EE M3}
\end{eqnarray}
The integral \eqref{EE M3} is UV divergent; therefore, we introduced a UV cutoff $U_{\rm max}$, as discussed above.
\begin{figure}[h]
    \centering
    \includegraphics[width=.5\textwidth]{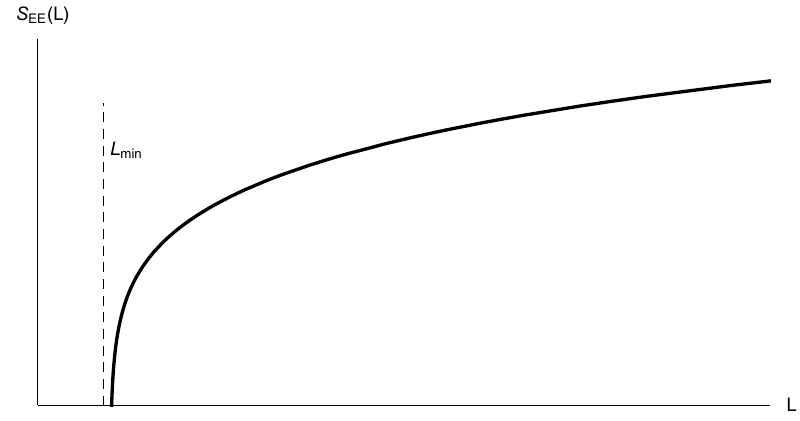}
    \caption{$S_{EE}(L)$.}
    \label{M3 T=0 EE S vs L}
\end{figure}
Evaluating the integral numerically, leads to the structure in figure (\ref{M3 T=0 EE S vs L}). 


\subsection{Large and small $L$ expansion of $S_{EE}(L)$}

In the last subsection we found an exact expression for the entanglement entropy $S_{EE}(L)$, \eqref{T=0 L}, \eqref{EE M3}. It is instructive to investigate the large and small $L$ behavior of $S_{EE}(L)$. The main motivations for this are to understand the dependence of the entropy on the UV cutoff $U_{\rm max}$, and for comparing to field theory results.

To study the behavior for large $L$, we define the variables
\begin{eqnarray}\label{rho Lhat}
\rho = \sqrt{k}U_0
\text{\; and\; } \hat{L} = \frac{L}{\sqrt{k}},
\end{eqnarray}
in terms of which one has
\begin{eqnarray}\label{Lhat}
\hat{L}(\rho)&=&\frac{2\sqrt{\alpha'}}{\rho} I(\rho) \ \qquad \ I(\rho)=\int_1^\infty dy \frac{\sqrt{\rho^2y^2+1}}{y^2\sqrt{y^2\left(\frac{\rho^2y^2+1}{\rho^2+1}\right)-1}}.
\end{eqnarray} 
Changing variables, $z=\rho y$, one has
\begin{eqnarray}\label{L(rho)}
\hat{L}(\rho)=2\sqrt{\alpha'}\int_\rho^\infty dz \frac{\sqrt{z^2+1}}{z^2\sqrt{\frac{z^2(z^2+1)}{\rho^2(\rho^2+1)}-1}}.
\end{eqnarray} 
We are interested in the large $L$, or equivalently small $U_0,\rho$ expansion of this integral. For $\rho>1$, one can write 
\begin{eqnarray}
\hat{L}(\rho)&=&\sqrt{\alpha'}\sum_{n=0}^\infty C_n\frac{(\rho^2+1)^{n+\frac{1}{2}}}{\rho^{2n+1}(2n+1)}~_2F_1\left[n,2n+1,2(n+1);-\frac{1}{\rho^2}\right],\label{L series}
\end{eqnarray}
where the coefficients $\{C_n\}$ are defined by
\begin{eqnarray}\label{Cn}
\frac{1}{\sqrt{1-x}}=\sum_{n=0}^{\infty}C_nx^n \text{ for } |x|<1.
\end{eqnarray} 
Expanding \eqref{L series} around $\rho=0$, one obtains
\begin{eqnarray}
\hat{L}(\rho)= \frac{2\sqrt{\alpha'}}{\rho}\left(1-\frac{\rho^4}{2}\log\rho+O(\rho^4)\right).\label{Lhat ads3}
\end{eqnarray}
The first term in the above expression is what one would obtain if the full background was $AdS_3$. The higher terms are due to corrections to the background. Inverting \eqref{Lhat ads3}, one gets (at large $L$)
\begin{eqnarray}
\rho=\frac{2\sqrt{\alpha'}}{\hat{L}}+16\frac{\alpha'^{5/2}}{\hat{L}^5}\log\frac{\hat{L}}{\sqrt{\alpha'}}~.\label{rho(Lhat)}
\end{eqnarray}

So far we discussed the form of \eqref{T=0 L} for large $L$, or equivalently small $U_0$. For large $U_0$ (\ie\ in the linear dilaton regime), $L$ behaves as 
\begin{eqnarray}\label{LD T=0 L}
L(U_0)&=& L_{\rm min}+\sqrt{\frac{\alpha'}{k}}\frac{1}{U_0^2}+O\left(\frac{1}{U_0^4}\right).
\end{eqnarray}

The entanglement entropy \eqref{EE M3} can be analyzed in a similar way. For small $U_0$, one finds
\begin{eqnarray}
S_{EE}&=&\left(\frac{\sqrt{k\alpha'}}{2G_N^{(3)}}\right)\left\{\frac{1}{2}(\rho_{\rm max}^2-\rho^2)+\log\left(\frac{\rho_{\rm max}}{\rho}\right)\right\} \nonumber \\
&& +\left(\frac{\sqrt{k\alpha'}}{2G_N^{(3)}}\right)\sum_{n=1}^{\infty} \frac{C_n(1+\rho^2)^n}{2\rho^{2(n-1)}(2n-1)}~_2F_1\left[n-1,2n-1,2n,-\frac{1}{\rho^2}\right].\label{EE M3 series}
\end{eqnarray}
where $\rho_{\rm max}=\sqrt{k}U_{\rm max}$. Expanding \eqref{EE M3 series} in $\rho$, and using \eqref{rho Lhat}, one finds
\begin{eqnarray}
S_{EE}&=&\frac{\sqrt{k\alpha'}}{2G_N^{(3)}}\left\{\frac{kU^2_{\rm max}}{2}+\log\left(\frac{2U_{\rm max}}{U_0}\right)-\frac{kU_0^2}{4}-\frac{3}{8}k^2U_0^4\log\left(U_0\right)+O(U_0^4)\right\}, \label{EE ads} 
\end{eqnarray}
Using \eqref{rho(Lhat)} we get
\begin{eqnarray}
S_{EE}=\frac{c}{3}\left\{\frac{\beta_H^2}{8\pi^2 L_{\Lambda}^2}+\log\left(\frac{L}{L_{\Lambda}}\right)-\frac{1}{4\pi^2}\left(\frac{\beta_H}{L}\right)^2+\frac{1}{8\pi^4}\left(\frac{\beta_H}{L}\right)^4\log\left(\frac{\beta_H}{L}\right)+O\left(\frac{\beta_H^4}{L^4}\right)\right\},\nonumber \\\label{SEE largeL}
\end{eqnarray}
where 
\begin{eqnarray}\label{Llambda}
 L_{\Lambda} = \frac{\sqrt{\alpha'}}{U_{\rm max}}
 \end{eqnarray} 
 is the UV cutoff, $\beta_H$ is the inverse Hagedorn temperature \eqref{H temp}, and $c$ the Brown -- Henneaux central charge 
 \begin{eqnarray}\label{c}
c=\frac{3\ell}{2G_N^{(3)}}=\frac{3\sqrt{k\alpha'}}{2G_N^{(3)}}.
\end{eqnarray}
In the linear dilaton regime (\ie\ $L\to L_{\rm min}$, $U_0\to\infty$), $S_{EE}$ behaves as 
\begin{eqnarray}\label{SEE LD exp}
S_{EE}&=& \frac{c}{6}\left\{\frac{\beta_H^2}{4\pi^2 L_{\Lambda}^2}+\log\left(\frac{\beta_H(L-L_{\rm min})}{L^2_{\Lambda}}\right)+O\left((L-L_{\rm min})^0\right) \right\}.
\end{eqnarray}
It would be interesting to understand better the radius of convergence of the large $L$ expansion \eqref{SEE largeL}, in analogy to what was done for correlation functions in \cite{Asrat:2017tzd}. We will leave this to future work.

\subsection{Casini-Huerta c-function}

\begin{figure}[h]
    \centering
    \includegraphics[width=.5\textwidth]{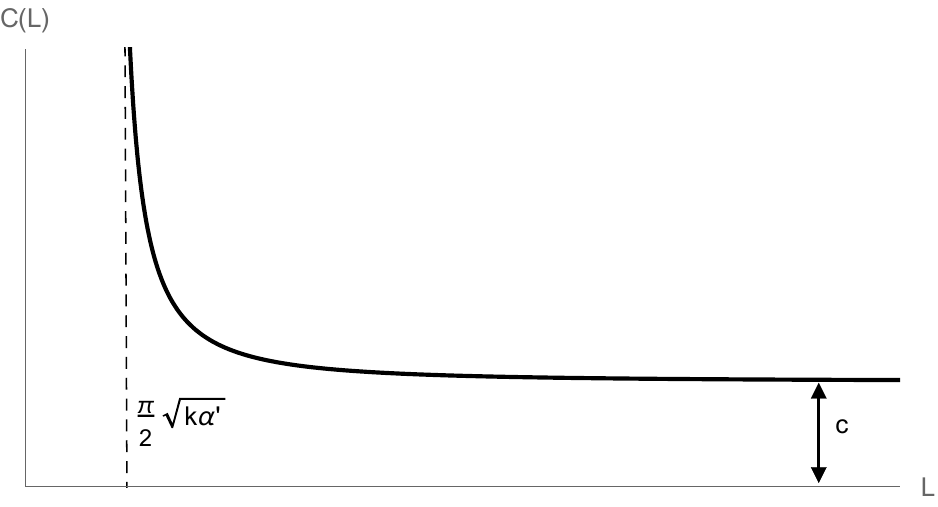}
    \caption{$C(L)$.}
    \label{M3 zero T EE C vs L}
\end{figure}

\begin{figure}[h]
    \centering
    \includegraphics[width=.5\textwidth]{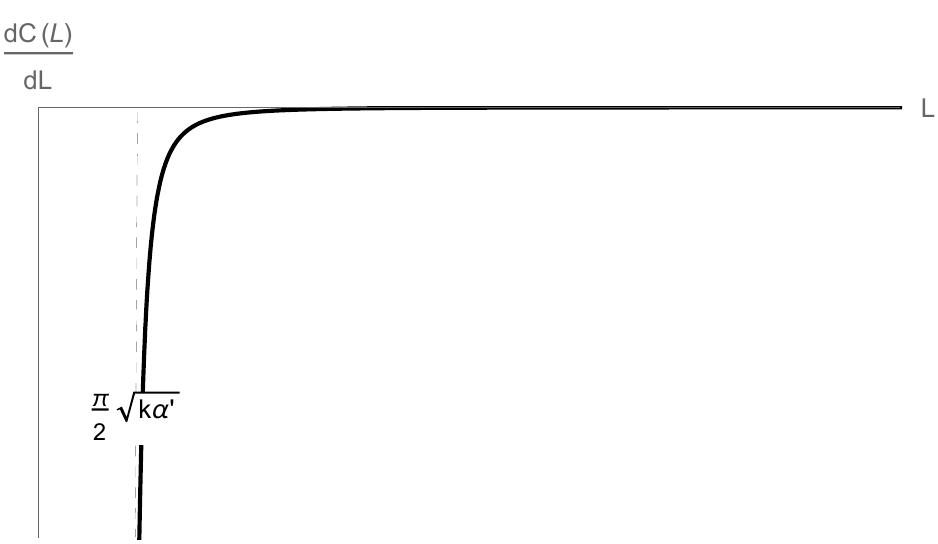}
    \caption{$C'(L)$.}
    \label{M3 zero T EE C' vs L}
\end{figure}

From the exact expression for the entanglement entropy, \eqref{T=0 L}, \eqref{EE M3}, we can compute the c-function \eqref{ccclll}. A numerical evaluation yields the results in figures (\ref{M3 zero T EE C vs L}), (\ref{M3 zero T EE C' vs L}). Note that:
\begin{enumerate}[(i)]
\item {While the entanglement entropy \eqref{EE M3} depends on the UV cutoff, $C(L)$ does not. As mentioned above, in a QFT governed by a UV fixed point this is necessarily the case, due to \eqref{see}, but here the theory is non-local in the UV, and this fact is non-trivial.}
\item {$C(L)\geq 0$, and is monotonically decreasing from the UV to the IR (\ie\ $C'(L)\leq 0$).  Thus, we see that the $c$-function can be generalized beyond the class of theories that approach a fixed point in the UV to the non-local theories discussed here, while preserving positivity and monotonicity.}
\end{enumerate}
 As $L\to\infty$ (\ie\ in the $AdS_3$ regime), $C(L)$ approaches a constant; $\lim_{L\to\infty} C(L)=c=6kN$. The large $L$ expansion of $C(L)$ is (inserting \eqref{SEE largeL} in \eqref{ccclll})
\begin{eqnarray}\label{CH fun ads}
C(L)&=&c\left\{1+\frac{1}{2\pi^2}\left(\frac{\beta_H}{L}\right)^2+\frac{1}{2\pi^4}\left(\frac{\beta_H}{L}\right)^4\log\left(\frac{L}{\beta_H}\right)+O\left(\left(\frac{\beta_H}{L}\right)^4\right)\right\}.
\end{eqnarray}
As $L\to L_{\rm min}$ (\ie\ in the linear dilaton regime), $C(L)$ diverges as 
\begin{eqnarray}\label{CH fun LD}
C(L)&=& \frac{c}{2}\left(\frac{L_{\rm min}}{L-L_{\rm min}}\right)+O\left((L-L_{\rm min})^0\right).
\end{eqnarray}
This divergence is due to the fact that LST has a Hagedorn density of states. Indeed, in \cite{Klebanov:2007ws} it was shown that in a theory whose spectrum contains a Hagedorn density of non-interacting states, the entanglement entropy diverges as $L$ approaches $\beta_H/2$ (see also \cite{Barbon:2008ut}). In our case, the divergence happens at $L_{\rm min}=\beta_H/4$. The difference
of a factor of two is likely due to the fact that here, unlike in \cite{Klebanov:2007ws}, the theory is strongly coupled. 

Another difference between our analysis here and that of  \cite{Klebanov:2007ws} is that there the
divergence was related to a jump of the entanglement entropy from order one to order $N^2$ (in a large $N$ gauge theory), or in the language of a holographic string dual from an entropy of order one to order $1/g_s^2$. Here, the divergence is visible in leading order in $g_s$. This can be understood from the point of view of the heuristic model mentioned at the end of section \ref{sec2.1}. It is due to the fact that we are dealing with a $T\bar T$ deformation of the block $\MM$ in a symmetric product $\MM^N/S_N$, so the analog of the discussion of 
\cite{Klebanov:2007ws} is visible to leading order in $1/N$.

\section{Field theory calculation}\label{sec4}

In this section we will discuss the entanglement entropy of an interval $I_L$ (see section \ref{sec2.2}) in $T\bar T$ deformed $CFT_2$. The main motivation  is to compare it to the string theory results of the previous section, particularly the large $L$ expansion \eqref{SEE largeL}, \eqref{CH fun ads}. 

We will consider a specific class of examples,\footnote{We expect the results of this section to depend only on the central charge of the original CFT, so one can consider any theory with the right central charge.} the theory of $N$ complex left and right-moving fermions, with $c=\bar c=N$. We start by reviewing the $CFT$ calculation in the formalism of \cite{Calabrese:2009qy}, and then calculate the leading correction to the CFT result \eqref{see} in the $T\bar T$ deformed $CFT_2$.

\subsection{Renyi entropy of $N$ complex fermions}

As reviewed in section \ref{sec2.2}, to calculate the entanglement entropy of a $CFT_2$ using the replica trick, we start with $n$ copies of the $CFT$, and construct the $\mathbb{Z}_n$ twist field $S_n$. In our case, the original $CFT$ contains the complex left-moving fermions $\psi^\alpha$, $\alpha=1,2,\cdots, N$, and their right-moving counterparts $\bar\psi^\alpha$, and taking $n$ copies of it leads to fermions $\psi^\alpha_i$, $\bar\psi^\alpha_i$, with $i$ the replica index, $i=1,2,\cdots, n$. The action of the $\mathbb{Z}_n$ generator  $\mathcal{T}$ on the fermions is\footnote{From here we discuss the left-moving sector only. There is an analogous story for the right-movers.}
\begin{eqnarray}\label{Zn action}
 \mathcal{T}:&&\psi^\alpha_i\to \psi^\alpha_{i+1}~ .
\end{eqnarray} 
The fermions satisfy the following ``periodicity condition'' in $i$: $\psi^\alpha_{i+n}=(-)^{n-1}\psi^\alpha_i$. 

After diagonalizing the action of $\mathbb{Z}_n$ on the fermions \eqref{Zn action} by a discrete Fourier transform,
\begin{eqnarray}\label{psi FT}
\tilde\psi_k^\alpha&=& \frac{1}{\sqrt{n}}\sum_{j=1}^{n}\psi_j^\alpha e^{2\pi i\frac{j(k-\frac{1}{2}(n-1))}{n}}, \qquad k=0,1,\cdots, n-1,
\end{eqnarray}
we find
\begin{eqnarray}\label{Zn act}
\mathcal{T}: \tilde{\psi}_k^\alpha\to \tilde{\psi}_k^\alpha e^{-\frac{2\pi i}{n}\left(k-\frac{1}{2}(n-1)\right)}.
\end{eqnarray}
The twist operator that implements the transformation \eqref{Zn act} for a particular $\alpha$ is  
\begin{eqnarray}\label{Snalpha}
S_n^\alpha=\prod_{k=0}^{n-1}s_k^\alpha,
\end{eqnarray}
where $s^\alpha_k$ is the twist field acting on $\tilde\psi_k^\alpha$. A simple way to construct it is to bosonize the fermions $\tilde\psi$, by writing them as 
\begin{eqnarray}\label{bose}
\tilde\psi_k^\alpha=e^{iH_k^\alpha},
\end{eqnarray}
where $H_k^\alpha$ is a canonically normalized scalar field, $\langle H(z)H(w)\rangle=-\ln(z-w)$. Then we have
\begin{eqnarray}\label{Spin}
s_k^\alpha=e^{{i\over n}(k-\half(n-1))H_k^\alpha}~.
\end{eqnarray}
The scaling dimension of $s_k^\alpha$ is
\begin{eqnarray}\label{dim skalpha}
\Delta(s_k^\alpha)=\frac{1}{2n^2}\left(k-\frac{(n-1)}{2}\right)^2.
\end{eqnarray}
Thus, the scaling dimension of the twist operator $S_n^\alpha$ \eqref{Snalpha} is given by
\begin{eqnarray}\label{dim Snalpha}
\Delta(S_n^\alpha)=\sum_{k=0}^{n-1}\Delta(s_k^\alpha)=\frac{1}{24}\left(n-\frac{1}{n}\right),
\end{eqnarray}
and for the total spin field, $S_n=\prod_{\alpha=1}^N S_n^\alpha$, we have
\begin{eqnarray}\label{dim Sn}
\Delta(S_n)=\Delta_n=\frac{c}{24}\left(n-\frac{1}{n}\right),
\end{eqnarray}
with $c=N$, in agreement with \eqref{dim twist}. 

\subsection{Order $\lambda$ correction to Renyi entropy}

We now turn on the $T\bar T$ deformation in the original $CFT$ of $N$ fermions, which corresponds to turning on the deformation 
\begin{eqnarray}\label{defff}
\delta\mathcal{L}=\lambda\sum_{l=1}^nT_l\bar{T}_l
\end{eqnarray}
in the product of $n$ $CFT$'s. To compute the entanglement entropy, we need to evaluate the two point function \eqref{trace} in the deformed theory \eqref{defff}. To first order in $\lambda$, we have  
\begin{eqnarray}\label{Pn}
\langle S_n(x)S_n(0)\rangle_\lambda=\frac{1}{|x|^{4\Delta_n}}-\lambda \sum_{l=1}^n\int d^2z\Big\langle T_l(z)\bar T_l(\bar z)S_n(x)S_n(0)\Big\rangle_0 + O(\lambda^2).
\end{eqnarray}
Conformal invariance of the theory with $\lambda=0$ implies that the three point function in \eqref{Pn} has the form 
\begin{eqnarray}\label{confinv}
\sum_{l=1}^n\Big\langle T_l(z)\bar T_l(\bar z)S_n(x)S_n(0)\Big\rangle_{\lambda=0} =\frac{C_n}{|z|^4|x-z|^4|x|^{4(\Delta_n-1)}}~.
\end{eqnarray}
where $C_n$ is a constant to be determined. 

To compute this constant, one rewrites the operator $\sum_{l=1}^nT_l\bar T_l$ in terms of the fermions $\psi$, $\bar\psi$,
\begin{eqnarray}\label{formt}
\sum_{l=1}^nT_l\bar T_l=\sum_{l=1}^n\sum_{\alpha,\beta=1}^N\psi^{*\alpha}_l\partial\psi^\alpha_l
\bar\psi^{*\beta}_l\bar\partial\bar\psi^\beta_l.
\end{eqnarray}
In terms of the Fourier transformed fermions $\tilde\psi$, \eqref{formt} takes the form 
\begin{eqnarray}\label{newformt}
\sum_{l=1}^nT_l\bar T_l=\sum_{k_1,\cdots k_4=1}^n\sum_{\alpha,\beta=1}^N\tilde\psi^{*\alpha}_{k_1}\partial\tilde\psi^\alpha_{k_2}
\bar{\tilde\psi}^{*\beta}_{k_3}\bar\partial\bar{\tilde\psi}^\beta_{k_4}\delta_{k_1-k_2+k_3-k_4,0}~.
\end{eqnarray}
Inserting \eqref{newformt} into \eqref{confinv}, we find that the only terms that contribute to the three point function are the ones with $k_1=k_2$ and $k_3=k_4$. Restricting to these terms, \eqref{newformt} takes the form $T_{\rm tot}\bar T_{\rm tot}$, where $T_{\rm tot}=\sum_l T_l$, and similarly for $\bar T$. Therefore, the constant $C_n$ in \eqref{confinv} can be determined from the stress tensor Ward identity to be 
\begin{eqnarray}\label{formcn}
C_n=\Delta_n^2~,
\end{eqnarray}
with $\Delta_n$ given by \eqref{dim Sn}.

Plugging \eqref{confinv}, \eqref{formcn} into \eqref{Pn}, we find that the order $\lambda$ correction to the two point function of $S_n$ is proportional to 
\begin{eqnarray}\label{integr}
\sim \int d^2 z \frac{\Delta_n^2}{|z-x|^4|z|^4|x|^{4\Delta_n-4}} \sim \frac{\Delta_n^2}{|x|^{4\Delta_n+2}}\ln(|x|\Lambda),
\end{eqnarray}
where $\Lambda$ is the UV cutoff, and we omitted an overall multiplicative constant.

The result \eqref{integr} together with \eqref{Pn} determines the Renyi entropy \eqref{renyi}, \eqref{trace} to order $\lambda$. To calculate the entanglement entropy $S_{EE}$, we need to divide by $n-1$ and take the limit $n\to 1$ \eqref{EntEntp}. Since $\Delta_n$ goes to zero like $n-1$, the order $\lambda$ contribution to $S_{EE}$ vanishes. 

It is interesting to compare the field theory analysis to the string theory results of the previous section, \eqref{SEE largeL}. In the field theory analysis, the expansion parameter is  $\lambda/|x|^2$, which is the same as the string theory expansion parameter $(\beta_H/L)^2$. Thus, in the notation of section \ref{sec3}, we found in the field theory analysis that  while the Renyi entropy receives a correction of the form $(\beta_H/L)^2\ln L$, the coefficient of this term in the entanglement entropy vanishes. This term is also absent in the expansion of the string theory result \eqref{SEE largeL}.

Of course, it is important to stress that these two calculations are done in different theories, as reviewed earlier in the paper. However, as we also mentioned, the two theories are closely related, and the calculation of the entanglement entropy provides one more example of this. It would be interesting to generalize the string theory calculation to that of the Renyi entropy (\eg\ using the results of \cite{Dong:2016fnf}), and check whether the agreement of the coefficient of this term persists for that case. 

In the string theory analysis \eqref{SEE largeL}, we found that while the $(\beta_H/L)^2\ln L$  term in the expansion was absent, there was a contribution of order $(\beta_H/L)^2$. It is natural to ask what is the origin of this contribution in the field theory expansion. One way it can arise is from contact terms. When evaluating \eqref{Pn} we used the three point function at separated points \eqref{confinv}. However, the integral over $z$ in \eqref{Pn} is also sensitive to contact terms. For example, we can introduce the contact term 
\begin{eqnarray}\label{conterm}
\sum_{l=1}^nT_l(z)\bar T_l(\bar z)S_n(0)=A_n\delta^2(z)\partial\bar\partial S_n(0).
\end{eqnarray}
Adding this to the calculation \eqref{Pn} leads to a term that goes like $(\beta_H/L)^2$ in \eqref{SEE largeL}, with a coefficient that depends on the behavior of $A_n$ as $n\to 1$. The coefficient $A_n$ is not determined by standard CFT data. It has to do with the geometry of the space of field theories (see \eg\ \cite{Kutasov:1988xb}), and can be chosen at will. In particular, we can choose it to reproduce the result of the string theory calculation.\footnote{In the bulk, the ambiguity related to the contact term (4.17)
     corresponds to the ambiguity of redefining the UV cutoff;
     the two terms that go like $\beta_H^2$ in (3.15) can mix 
     by replacing the UV cutoff $U_{\rm{max}}^2$ with 
     $U_{\rm{max}}^2+{\rm{constant}\over L^2}$.}

\section{Bulk calculation: finite temperature}\label{sec5}

In this section we generalize the analysis of section \ref{sec3} to finite temperature. To do this we replace the geometry \eqref{background} by a black hole in $\mathcal{M}_3$,
\begin{eqnarray}
ds^2&=&-\frac{f_1}{f}dt^2+\frac{1}{f}dx^2+k\alpha'f_1^{-1}\frac{dU^2}{U^2},\nonumber \\
e^{2\Phi}&=&\frac{g_s^2}{kU^2}f^{-1},\label{background T}\\
dB&=&\frac{2i}{U^2}f^{-1}\epsilon_3~ ,\nonumber
\end{eqnarray}
where $f=1+\frac{1}{kU^2}$ and $f_1=1-\frac{U_T^2}{U^2}$. The location of the horizon, $U_T$, is related to the temperature via:
\begin{eqnarray}\label{BH temp}
T=\frac{1}{2\pi}\sqrt{\frac{U_T^2}{\alpha'(1+kU_T^2)}}.
\end{eqnarray}
When $\sqrt{k}U_T\ll 1$, the horizon is deep inside the $AdS_3$ region in $\MM_3$, and is described to a good approximation by the BTZ black hole.  When $\sqrt{k}U_T\gg 1$, the horizon is deep inside the linear dilaton regime and the solution is  described by the coset $\frac{SL(2,\IR)\times U(1)}{U(1)}$ (see \eg\ \cite{Giveon:2005mi}).

\subsection{Holographic entanglement entropy}

\begin{figure}[h]
    \centering
    \includegraphics[width=.5\textwidth]{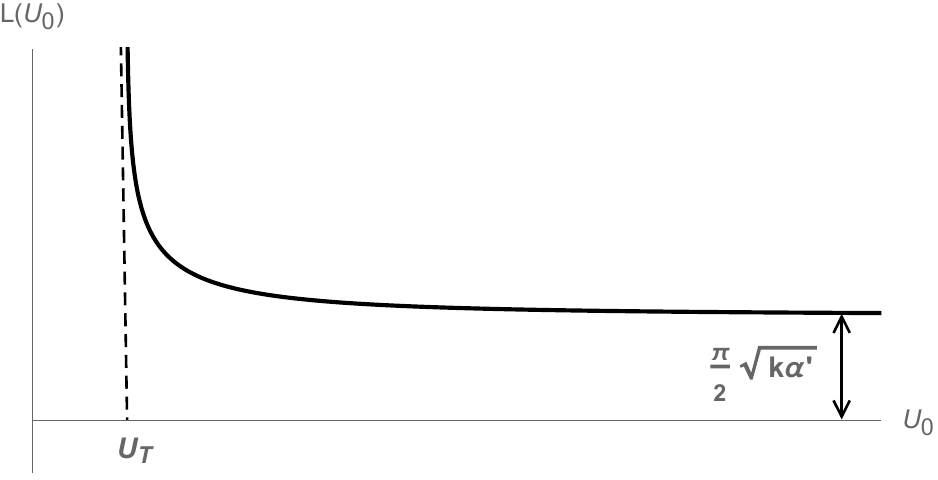}
    \caption{The size of the entangling region $L$ as a function of $U_0$ in the background \eqref{background T}.}
    \label{M3 finite T EE L vs U0}
\end{figure}

Writing down the area functional \eqref{EE} for a surface $U=U(x)$ in the geometry \eqref{background T} gives rise to the equation of motion
\begin{eqnarray}\label{config}
\frac{U^2\sqrt{U^2-U_T^2}\sqrt{kU^2+1}}{\sqrt{U^2(U^2-U_T^2)+\alpha'(kU^2+1)(\partial_x U)^2}}&=& U_0\sqrt{kU_0^2+1}~,
\end{eqnarray}
with the initial condition that $U(x=0)=U_0>U_T$ and $\partial_x U|_{x=0}=0$. As in the zero temperature case, \eqref{T=0 L}, the size of the line segment on the boundary, $L$, is related to $U_0$ via
\begin{eqnarray}\label{L T}
L(U_0)=\frac{2\sqrt{\alpha'}}{U_0}\int_1^\infty dy \frac{\sqrt{kU_0^2y^2+1}}{\sqrt{y^2\left(y^2-\frac{U_T^2}{U_0^2}\right)}\sqrt{y^2\left(\frac{kU_0^2y^2+1}{kU_0^2+1}\right)-1}}~.
\end{eqnarray}
The integral is again convergent in the UV and one does not need to introduce a UV cutoff.
Figure (\ref{M3 finite T EE L vs U0}) shows a numerical plot of $L$ as a function of $U_0$. The profile does not change much as we take $U_T$ continuously from the $AdS_3$ region to the linear dilaton one. The IR and UV behaviors of $L(U_0)$ are similar to those found for $T=0$ in section \ref{sec3}, except now $U_0$ is bounded from below by $U_T$.

 The entanglement entropy is given by
\begin{eqnarray}
S_{EE} &=& \frac{\sqrt{k\alpha'}}{2G_N^{(3)}}\int_{U_0}^{U_{\rm max}} \frac{dU}{U} \frac{U^2(kU^2+1)}{\sqrt{U^2\left(U^2-U_T^2\right)}\sqrt{1-\left(\frac{kU_0^2+1}{kU^2+1}\right)\frac{U_0^2}{U^2}}}~.\label{SEEC T}
\end{eqnarray}

\begin{figure}[h]
    \centering
    \includegraphics[width=.5\textwidth]{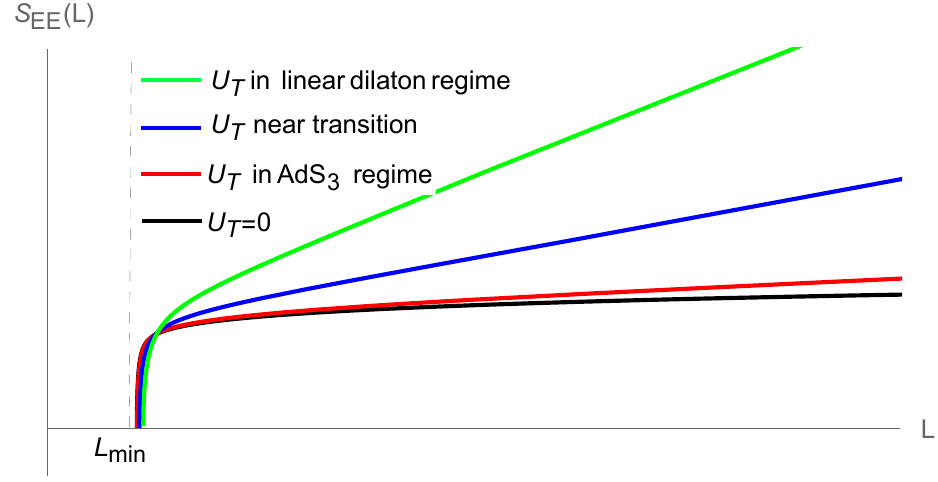}
    \caption{The entanglement entropy $S_{EE}$ as a function of $L$ in $\mathcal{M}_3$ at finite temperature for different horizon radii ranging from the $AdS_3$ regime to the linear dilaton regime.}
    \label{M3 finite T EE S vs L different uT}
\end{figure}
Figure (\ref{M3 finite T EE S vs L different uT}) shows a numerical plot of $S_{EE}$ as a function of $L$ for different values of $U_T$.

\subsection{Large and small $L$ expansion of entanglement entropy}

As in the zero temperature case, we can investigate the large and small $L$ behavior of the entanglement entropy in the background \eqref{background T}.  For large $U_0$, $L$ behaves as 
\begin{eqnarray}\label{L LD reg}
L(U_0)= \frac{\pi}{2}\sqrt{k\alpha'}+\sqrt{\frac{\alpha'}{k}}\frac{(2+kU_T^2)}{2U_0^2}+O\left(\frac{1}{U_0^4}\right).
\end{eqnarray}
As expected, in the limit $U_0\to\infty$, $L$ approaches the same constant value ($L=L_{min}=\frac{\pi}{2}\sqrt{k\alpha'}$) as in the case of zero temperature. 

When the horizon is deep inside the $AdS_3$ region, and the RT surface is hanging deep in the bulk, one gets
\begin{eqnarray}\label{L ads reg}
L&=&\frac{2\sqrt{\alpha'}}{U_T}\tanh^{-1}\left(\frac{U_T}{U_0}\right)+\text{sub-leading terms}.
\end{eqnarray}

When the horizon is deep inside the $AdS_3$ regime, with $U_0$ very close to the horizon, the entanglement entropy takes the form
\begin{eqnarray}
S_{EE} &\sim&\frac{c}{3}\left\{\frac{\beta_H^2}{8\pi^2 L_{\Lambda}^2}+  \log\left[\frac{\beta_H}{L_\Lambda}\sinh\left(\frac{\pi L\rho_T}{\beta_H}\right)\right] \right\},  \label{SC ads reg}
\end{eqnarray}
where  $\rho_T=\sqrt{k}U_T$. To obtain this result, we have used the fact that $\sqrt{k}U_{\rm max}\gg \rho_T,\rho_0$. 

In the limit $U_0\to \infty$, $S_{EE}$ behaves as
\begin{eqnarray}\label{SC LD reg}
S_{EE}
&=&\frac{c}{6}\left\{\frac{\beta_H^2}{4\pi^2 L_{\Lambda}^2}+\frac{(2+\rho_T^2)}{2}\log\left(\frac{\beta_H}{L_{\Lambda}^2}(L-L_{\rm min})\right)+O\left((L-L_{\rm min})^0\right) \right\}.
\end{eqnarray}

\section{Discussion}\label{sec6}

The main goal of this note was to study the entanglement entropy of an interval of length $L$ in the models discussed in \cite{Giveon:2017nie,Giveon:2017myj,Asrat:2017tzd}. In particular, we were interested in the consequences of non-locality of these models for this observable, and in comparing its properties to that of the closely related $T\bar T$ deformed $CFT_2$. 

We found that the entropic $c$-function \eqref{ccclll} has the following properties in these models:
\begin{enumerate}[(i)]
\item {It is independent of the UV cutoff, just like in local QFT.}
\item {It is monotonically decreasing along the RG (\ie\ $C'(L)<0$), just like in local QFT.}
\item {Unlike local QFT, where the $c$-function goes to a constant in the UV limit $L\to 0$ (equal to the central charge of the UV fixed point), in the models of \cite{Giveon:2017nie,Giveon:2017myj,Asrat:2017tzd} the $c$-function is not bounded from above; in fact it diverges as $L\to L_{\rm min}$, where $L_{\rm min}=\frac{\pi}{2}\sqrt{k\alpha'}$ (see figure (\ref{M3 zero T EE C vs L})).}
\item {The large $L$ expansion of the entanglement entropy has a similar structure to that of  $T\bar T$ deformed $CFT_2$.}
\end{enumerate}

While our results showed agreement between the string theory analysis of the models of \cite{Giveon:2017nie,Giveon:2017myj,Asrat:2017tzd} and $T\bar T$ deformed $CFT_2$, we would like to mention a case where the two seem to disagree. 
In the recent paper \cite{Cardy:2018sdv}, Cardy discussed the entanglement entropy of a half line in a massive $QFT_2$ with mass parameter $m$ in the presence of a $\lambda T\bar T$ deformation.\footnote{After the publication of the original version of the current paper, \cite{Cardy:2018sdv} was revised. The new versions do not contain this result.} His result, to first order in the dimensionless parameter $\lambda m^2$, has the form 
\begin{eqnarray}\label{cardy}
S_{EE}= \frac{c}{6} \log\frac{1}{ma}-\frac{c^2\lambda m^2}{72\pi}(\log(am))^2+ O(\lambda^2m^4\log am).
\end{eqnarray}
One can perform the same calculation in the models of \cite{Giveon:2017nie,Giveon:2017myj,Asrat:2017tzd}. For $\lambda=0$, this was discussed in \cite{Nishioka:2009un}. The mass parameter is introduced by taking the radial coordinate in $AdS_3$ to be bounded from below, $U\ge U_{\rm min}$. The entanglement entropy of the half line is obtained by calculating the length of the line labeled by $U$ at a given value of the spatial coordinate $x$. This gives 
\begin{eqnarray}\label{ads cutoff}
S_{EE}=\frac{c}{6}\ln \frac{U_{\rm max}}{U_{\rm min}}=\frac{c}{6}\ln \frac{\xi}{L_{\Lambda}},
\end{eqnarray}
where $\xi=\frac{\sqrt{\alpha'}}{U_{\rm min}}$ is the correlation length of the boundary field theory.

It is easy to generalize the calculation to non-zero $\lambda$, by calculating the length of the above line in $\MM_3$ rather than $AdS_3$. This gives
\begin{eqnarray}\label{M3 cutoff}
S_{EE}=\frac{1}{2}\left(\frac{\sqrt{k\alpha'}}{2G_N^{(3)}}\right)\int_{U_{\rm min}}^{U_{\rm max}}\frac{dU}{U}(kU^2+1)
=\frac{c}{6}\left\{ \log\frac{\xi}{L_\Lambda}+\frac{k\alpha'}{2}\left(\frac{1}{L_\Lambda^2}-\frac{1}{\xi^2}\right) \right\}.
\end{eqnarray}
Comparing this to the result of  \cite{Cardy:2018sdv}, \eqref{cardy}, with $m=1/\xi$, $a=L_\Lambda$, we see that the term that goes like $\lambda m^2\ln^2(am)$ in \eqref{cardy} is absent in \eqref{M3 cutoff}. It would be interesting to understand the origin of the discrepancy between the two calculations. 

It would also be interesting to generalize the calculation of section \ref{sec4} to higher orders in $\lambda$, and compare to the exact results of section \ref{sec3}. Another closely related calculation that would be interesting to do is that of Wilson lines in the geometry $\MM_3$.

\section*{Acknowledgements} 

We thank O. Aharony, J. Sonnenschein and M. Smolkin for discussions and M. Asrat and J. Kudler-Flam for pointing out an error in a previous version of this paper. The work of AG and NI is supported in part by the I-CORE Program of the Planning and Budgeting Committee and the Israel Science Foundation (Center No. 1937/12), and by a center of excellence supported by the Israel Science Foundation (grant number 1989/14). The work of DK is supported in part by DOE grant DE-SC0009924. DK thanks Tel Aviv University and the Hebrew University for hospitality during part of this work.

\end{document}